\documentclass{aa}
\usepackage{txfonts}
\usepackage{graphicx}
\usepackage{caption}
\usepackage{subcaption}
\usepackage{color}
\usepackage{tablefootnote}

\usepackage{tikz}
\usetikzlibrary{arrows,shapes,backgrounds}

\begin{document} 
\titlerunning{}
\authorrunning{L.Koutoulidis et al.}
\titlerunning{Host galaxy properties of X-ray AGN}

\title{Host galaxy properties of X-ray AGN in the Local Universe}

\author{L. Koutoulidis,
          \inst{1}
          G. Mountrichas,
          \inst{2}
          I. Georgantopoulos,
          \inst{1}
         E. Pouliasis,
	 \inst{1}
         M. Plionis
          \inst{1}}
\institute{IAASARS, National Observatory of Athens, I. Metaxa \& V. Pavlou 1, Penteli, 15236, Greece
\and 
	Instituto de Fisica de Cantabria (CSIC-Universidad de Cantabria), Avenida de los Castros, 39005 Santander, Spain}

\abstract {We study the host galaxy properties of active galactic nuclei (AGN) that have been detected in X-rays in the nearby Universe ($\rm z<0.2$). For that purpose, we use the catalogue provided by the ROSAT-2RXS in the 0.1-2.4\,keV energy band, one of the largest X-ray datasets with spectroscopic observations. Our sample consists of $\sim 900$ X-ray AGN. The catalogue provides classification of the sources into type 1 and 2, based on optical spectra. $\sim 25\%$ of the AGN are type 2. We use the available optical, near-IR and mid-IR photometry to construct SEDs. We measure the stellar mass ($M_*$) and star formation rate (SFR) of the AGN, by fitting these SEDs with the X-CIGALE code. We compare the $M_*$ and SFR of the two AGN populations, taking into account their different redshift and luminosity distributions. Based on our results, type 2 AGN tend to live in more massive galaxies compared to their type 1 counterparts ($\rm log\,[M_*(M_\odot)]=10.49^{+0.16}_{-0.10}$ vs. $10.23^{+0.05}_{-0.08}$), in agreement with previous studies at higher redshifts. In terms of SFR, our analysis shows, that in the nearby Universe, the number of X-ray AGN that live in quiescent systems is increased compared to that at higher redshifts, in accordance with previous studies in the local universe. However, the majority of AGN ($\sim 75\%$) live inside or above the main sequence.}

\keywords{active galaxies --
                X-rays--
                galaxies}
   
\maketitle

\section{Introduction}

Supermassive Black Holes (SMBH), with masses $\rm M_{BH}>10^6\,M_\odot$ are placed at the centre of most, if not all, galaxies including our own. Both observations and theoretical models have shown that there is a linear relation between the growth of the SMBH and the host galaxy evolution  \citep{Magorrian1998,Ferrarese2000,Gebhardt2000,Haring2004,Kormendy2013}. However, the details of the physical mechanism that connects them is still not well understood. 

The growth of the SMBH occurs when matter falls in the inner region of the galaxy, at the proximity of the SMBH. As it grows the SMBH releases huge amounts of energy across the entire electromagnetic spectrum. In such cases, the inactive galaxy turns into an Active Galactic Nucleus \citep[AGN; e.g.,][and references therein]{Hick2018}.

X-rays consist of one of the most reliable methods of detecting AGN since a large part of the AGN population, the obscured subsample, although detected in  X-rays is not detected in the optical regime of the spectrum \citep[][and references therein]{Hasinger2008,Hick2018}. Obscured AGN represent a considerable fraction of the total AGN population \citep[e.g.][]{Akylas2006, Ueda2014}. Therefore, studying the origin and nature of obscured AGN constitutes a significant challenge to reveal the complete AGN population. Some studies claim that different levels of obscuration correspond to different stages of the SMBH growth \citep[evolutionary model; e.g., ][]{Ciotti1997, Hopkins2006, Hopkins2008a, Somerville2008}. Those observations contradict geometric models such as the unification model in which the observed differences among type 1 and type 2 properties depend on the viewing angle that the AGN is observed \citep{Antonucci1993, Urry1995, Netzer2015}.

One approach to examine whether type 1 and 2 AGN constitute the same or different populations is to examine the properties of their host galaxies. There are two main parameters that are related to the early stages of the formation and evolution of galaxies, namely their stellar mass and star formation rate \citep{Kauffmann2003}. Previous studies have used different criteria to classify AGN (e.g. X-ray criteria based on hardness ratio or the hydrogen column density, optical spectra, optical/mid-IR colours). 

The studies of \cite{Mountrichas2021a} and \cite{Masoura2021} analyzed X-ray AGN observed by the Chandra X-ray Observatory within the 9.3 deg$^2$ Bo$\rm \ddot{o}$tes field of the NDWFS and X-ray AGN in the XXL-North respectively. They concluded that X-ray absorption is not linked with the properties of the galaxy. \cite{Lanzuisi2017}, used X-ray AGN in the COSMOS field and found that, although, obscured and unobscured AGN have similar star formation rate (SFR), unobsured AGN tend to have lower stellar masses than their obscured counterparts. \cite{Suh2019} examined the effects of the nuclear activity on the star formation in both type 1 and type 2 AGN host galaxies in the COSMOS field and found that it is not conclusive whether AGN activity plays a role in quenching the star formation in galaxies. They proposed that stellar mass might be the primary factor related to suppressing both star formation and AGN activity.

Recent studies that used optical criteria to classify their sources, found that both populations have similar SFR, but type 2 AGN tend to reside in more massive galaxies than their type 1 counterparts \citep[e.g.][]{Zou2019, Mountrichas2021a}. \cite{Shimizu2015}, used 122 ultra-hard X-ray AGN at $\rm z<0.05$ from the {\it{Swift}} Burst Alert Telescope \citep{Bat2005} and found similar SFR for the two AGN types, but their results suggest that, in the local universe, a high fraction of both type 1 and 2 AGN live in quiescent systems.

In the present work, we utilize a large, spectroscopic X-ray sample,  provided by the SPIDERS-2RXS survey, to examine the host galaxy properties of type 1 and 2 AGN. Their classification is based on optical spectra. This catalogue provides the largest systematic spectroscopic observations of an X-ray selected sample, making it an ideal laboratory to study AGN and their host galaxies. We construct the spectral energy distribution (SEDs) of the sources, using optical, near-IR and mid-IR photometry and use the X-CIGALE code to fit these SEDs. Our goal is to examine the host galaxy properties of the two AGN types and compare our findings with those at higher redshifts \citep[e.g.][]{Zou2019, Mountrichas2021a} and in the local Universe \citep[e.g.][]{Shimizu2015} that have also classified their AGN using optical criteria.

\begin{table*}
\caption{The models and the values for their free parameters used by X-CIGALE for the SED fitting of our galaxy sample. } 
\centering
\setlength{\tabcolsep}{1.mm}
\begin{tabular}{cc}
       \hline
Parameter &  Model/values \\
	\hline
\multicolumn{2}{c}{Star formation history: delayed model and recent burst} \\
Age of the main population & 1000, 5000, 7000, 9000, 10000, 11000, 12000 Myr \\
e-folding time & 1000, 3000, 5000, 7000, 9000, 10000, 11000, 12000 Myr \\ 
Age of the burst & 50 Myr \\
Burst stellar mass fraction & 0.0, 0.005, 0.01, 0.015, 0.02, 0.05, 0.10, 0.15, 0.18, 0.20 \\
\hline
\multicolumn{2}{c}{Simple Stellar population: Bruzual \& Charlot (2003)} \\
Initial Mass Function & Salpeter\\
Metallicity & 0.02 (Solar) \\
\hline
\multicolumn{2}{c}{Galactic dust extinction} \\
Dust attenuation law & Calzetti et al. (2000) \\
Reddening $E(B-V)$ & 0.0, 0.1, 0.2, 0.3, 0.4, 0.5, 0.6, 0.7, 0.8, 0.9 \\ 
\hline
\multicolumn{2}{c}{Galactic dust emission: Dale et al. (2014)} \\
$\alpha$ slope in $dM_{dust}\propto U^{-\alpha}dU$ & 1.0, 1.5, 2.0, 2.5, 3.0 \\
\hline
\multicolumn{2}{c}{AGN module: SKIRTOR)} \\
Torus optical depth at 9.7 microns $\tau _{9.7}$ & 3.0, 7.0 \\
Torus density radial parameter p ($\rho \propto r^{-p}e^{-q|cos(\theta)|}$) & 1.0 \\
Torus density angular parameter q ($\rho \propto r^{-p}e^{-q|cos(\theta)|}$) & 1.0 \\
Angle between the equatorial plan and edge of the torus & $40^{\circ}$ \\
Ratio of the maximum to minimum radii of the torus & 20 \\
Viewing angle  & $30^{\circ}\,\,\rm{(type\,\,1)},70^{\circ}\,\,\rm{(type\,\,2)}$ \\
AGN fraction & 0.0, 0.01, 0.1, 0.2, 0.3, 0.4, 0.5, 0.6, 0.7, 0.8, 0.9, 0.99 \\
Extinction law of polar dust & SMC \\
$E(B-V)$ of polar dust & 0.0, 0.2, 0.4 \\
Temperature of polar dust (K) & 100 \\
Emissivity of polar dust & 1.6 \\
\hline
\label{table_cigale}
\end{tabular}
\tablefoot{For the definition of the various parameter see section \ref{sec_cigale}.}
\end{table*}

\section{Data}
\label{sec_data}

In this study, we use the DR16-SPIDERS-2RXS catalogue. The catalogue is fully described in  \cite{Comparat2020}. In brief, this dataset utilizes the second ROSAT all sky survey source catalogue (hereafter 2RXS) which contains approximately 135000 X-ray detections in the $0.1-2.4$ keV energy band, with a likelihood threshold of $6.5$ \citep{Boller2016}. A novel Bayesian statistics based algorithm \citep[NWAY;][]{Salvato2018} was used to find reliable counterparts at other wavelengths. Then, the SDSS spectroscopic information is added, based on the optical position in the SPIDERS-DR16 footprint \citep{Dwelly2017}. This results in 19821 X-ray sources, 10336 of which are AGN, that cover an area of $5128\,\rm deg^2$ with flux limit of $\rm f_x=10^{-12.5}\,erg\,s^{-1}\,cm^{-2}$. Additionally to the optical (SDSS) photometry, sources also have near-IR from 2MASS \citep{Mass2006} and mid-IR photometry from allWISE \citep{Wright2010} datasets.

Following \cite{Comparat2020}, we selected objects with reliable redshifts $(CONFBEST==3)$ excluding $‘STARS’,’GALAXIES’ and ‘QSO’$ that belong to cluster of galaxies.
According to \cite{Boller2016} a detection likelihood threshold $>6.5$ leads to a large fraction of spurious sources ($\sim 30\%$). Thus, we adopted a more strict threshold for the detection criterion, specifically $exiML >10$. To focus in the nearby Universe, we selected sources with spectroscopic redshifts $\rm z<0.2$. We also restricted our sample to those AGN with X-ray luminosity, L$\rm _X>10^{42}\,erg\,s^{-1}$, to avoid contamination by non AGN systems. Applying the above criteria, our sample consists of 1021 sources.

To classify sources as unobscured/type 1 (broad emission lines) and obscured/type 2 (narrow emission lines) AGN,  we used  the spectral classification provided in \cite{Comparat2020}. We considered as type 1 sources with spectroscopic classes $ ‘BLAGN’, ‘QSO’,’BALQSO’,’QSO-BAL’$. As type 2 AGN we included sources with spectroscopic class $‘NLAGN’$. Our sample consists of 760 type 1 and 255 type 2 AGN, at $\rm z<0.2$.

\section{Analysis}

In this section, we describe the SED fitting process we performed to estimate the galaxy properties (SFR and stellar mass) and present the requirements we applied to include only sources with the most robust measurements, in our analysis.

\subsection{X-CIGALE}
\label{sec_cigale}

To calculate the properties of the galaxies that host the X-ray AGN of the sample used in this work, we perform SED fitting. For that purpose, we use the X-CIGALE code \citep{Yang2020} that is a new branch of the CIGALE fitting algorithm \citep{Boquien2019} and has been used widely in the literature
 \citep[e.g.][]{Pouliasis2020}. X-CIGALE can model the X-ray emission of galaxies and the presence of polar dust that accounts for extinction of the UV and optical emission in the poles of AGN \citep{Yang2020, Mountrichas2021a}. In our analysis, we do not use the X-ray flux, f$_X$, of the AGN in the fitting process. X-CIGALE requires the intrinsic, that is the X-ray absorption corrected, X-ray flux of the source. The X-ray absorption is known for $\sim73\%$ of our sources \citep{Boller2016}. Thus, including the f$_X$ in our SED fitting process would reduce the size of our sample whilst it would not affect the stellar mass and SFR measurements \citep{Mountrichas2021a}. Alternatively, observed, but hard (e.g. 2-10\,keV) X-ray fluxes can be included in the fitting analysis \citep{Yang2020}, since at these energies the X-ray flux is only mildly affected by the X-ray absorption. However, these energy bands are not available for our dataset.

A delayed star formation history (SFH) model, with the functional form SFR$\propto t\times \exp(-t/{\tau})$, is applied to built the galaxy component. The model also includes a star formation burst, in the form of a constant ongoing star formation that is not allowed to be longer than 50\,Myr. Stellar emission is modelled using the \cite{Bruzual_Charlot2003} single stellar populations template with the initial mass function (IMF) of Salpeter and metallicity equal to the solar value (0.02). Stellar emission is attenuated following \cite{Charlot_Fall_2000}. The IR SED of the dust heated by stars is implemented with the \cite{Dale2014} model, without the AGN component. AGN emission is modelled using the SKIRTOR templates \citep{Stalevski2012, Stalevski2016}. SKIRTOR is a clumpy two-phase torus model that considers an anisotropic, but constant, disk emission. A detailed description of the SKIRTOR implementation in (X-) CIGALE is given in \cite{Yang2020}. The AGN fraction, $\rm frac_{AGN}$, is defined as the ratio of the AGN IR emission to the total IR emission of the galaxy. A polar dust component ($E_{B-V}$) is added and modelled as a dust screen absorption and a grey-body emission. The Small Magellanic Cloud extinction curve \citep[SMC;][]{Prevot1984} is adopted. Re-emitted grey-body dust is parameterized with a temperature of $\rm 100\,K$ and emissivity index of 1.6. The modules and input parameters we use in our analysis are presented in Table \ref{table_cigale}.

In our analysis, we do not fix the inclination angle to a value that corresponds to the optical classification of the AGN. The addition of polar dust in the fitting process makes the definition of obscured and unobscured AGN more complex. The UV/optical SED of type 2 AGN is not affected, since it is already absorbed by the dusty torus. However, polar dust reddens the UV/optical SED of type 1 AGN. For a detailed discussion on the effect of polar dust on the AGN classification by X-CIGALE and its comparison with the optical classification, see section 5.1 of \cite{Mountrichas2021a}. Their analysis also showed that the misclassification of some sources by X-CIGALE, does not affect the reliability of the host galaxy measurements. The modules and input parameters we use in our analysis are presented in Table \ref{table_cigale}.

\subsection{Exclusion of sources with non reliable measurements}
\label{sec_bad_fits}

\begin{table}
\caption{Number of X-ray AGN used in the different parts of our analysis (see text for more details).}
\centering
\setlength{\tabcolsep}{3.mm}
\begin{tabular}{cccc}
  & stellar mass & SFR & SFR$_{norm}$ \\
\hline
 Total AGN & 944 & 860 &859\\
 type 1 & 729 & 673 & 672\\
 type 2 & 215 & 187 & 187\\
 \hline
\label{table_numbers}
\end{tabular}
\end{table}

It is important in our analysis to only keep sources for which the SFR and stellar mass estimates are reliable. Towards this end, we first exclude sources that their fit has reduced $\chi ^2$, $\chi ^2_{red}>5$. This criterion is based on visual inspection of the SED fits and has been adopted in previous studies \citep[e.g.,][]{Masoura2018, Mountrichas2019, Buat2021}. Increasing the limit to  $\chi ^2_{red}>6$,  adds more sources in our sample but most of them with bad fits and thus unreliable host galaxy measurements. Reducing the threshold to  $\chi ^2_{red}>4$, would exclude additional sources, the vast majority of which have reliable fits. This criterion eliminates 133  AGN (9\% of the sample). For each parameter calculated by the SED fitting process, X-CIGALE estimates two values. One is evaluated from the best-fit model and one that weighs all models allowed by the parametric grid, with the best-fit model having the heaviest weight \citep{Boquien2019}. This weight is based on the likelihood, exp ($-\chi^2/2)$, associated with each model. A large difference between these two values for a specific parameter, indicates that the fitting process did not result in a reliable estimation for this parameter \citep{Mountrichas2021b, Mountrichas2021c, Buat2021}. Thus, when we compare the stellar masses of type 1 and 2 AGN, we only include in our analysis sources with $\rm \frac{1}{5}\leq \frac{M_{*, best}}{M_{*, bayes}} \leq 5$, where  M$\rm _{*, best}$ and M$\rm _{*, bayes}$ are the best and Bayesian fit values of M$_*$, respectively. This method to exclude unreliable estimations have been applied in recent studies \citep{Mountrichas2021b, Mountrichas2021c, Buat2021}. Using different values for the boundaries of the criterion, i.e., 0.1-0.33 for the lower limit and 3-10 for the upper limit does not affect the results of our analysis. This criterion reduces the sample to 944 X-ray AGN (93\% of the initial dataset). 729 of these systems are type 1 and 215 are type 2 (Table \ref{table_numbers}). Similarly, for the comparison of the SFR of the two AGN types, we require $\rm \frac{1}{5}\leq \frac{SFR_{best}}{SFR_{bayes}} \leq 5$, where SFR$\rm _{best}$ and SFR$\rm _{bayes}$ are the best and Bayesian values of SFR, respectively, estimated by X-CIGALE. This reduces our sample to 860 X-ray AGN (85\% of the initial dataset). 673 of these sources are type 1 and 187 are type 2. We note that throughout our analysis, we use the bayes calculations of X-CIGALE for the various parameters.

In {Fig. \ref{fig_SEDs}, we present examples of SED from sources used in our analysis. Top panels show AGN classified as type 1, based on their optical spectra. Bottom panels, present AGN with an optical classification of type 2.  Fig. \ref{fig_stacked} presents the stacked SEDs at rest-frame for the AGN samples of type 1 (blue) and type 2 (green). The median SEDs of the total emission (host galaxy and AGN components) are plotted with solid lines, while the shaded areas correspond to 15$^{th}$ up to 85$^{th}$ percentiles at each wavelength. The lower panel shows the ratio of the host dust and stellar attenuated emission of the two populations. This plot is further discussed in section \ref{sec_results}.

\begin{figure*}
\centering
\begin{subfigure}{.48\textwidth}
  \centering
  \includegraphics[width=1.0\linewidth, height=7.cm]{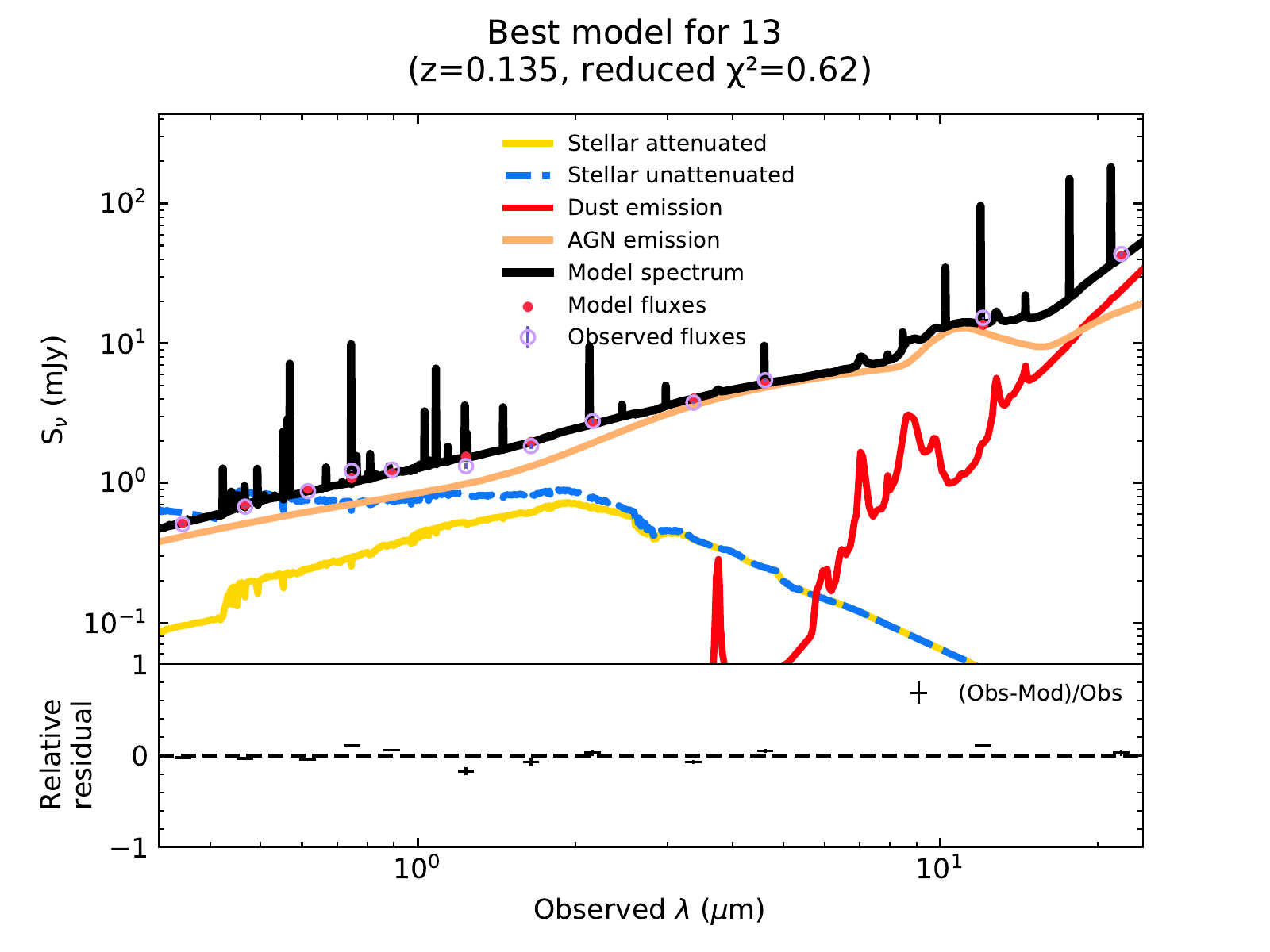}
  \label{}
\end{subfigure}
\begin{subfigure}{.48\textwidth}
  \centering
  \includegraphics[width=1.0\linewidth, height=7.cm]{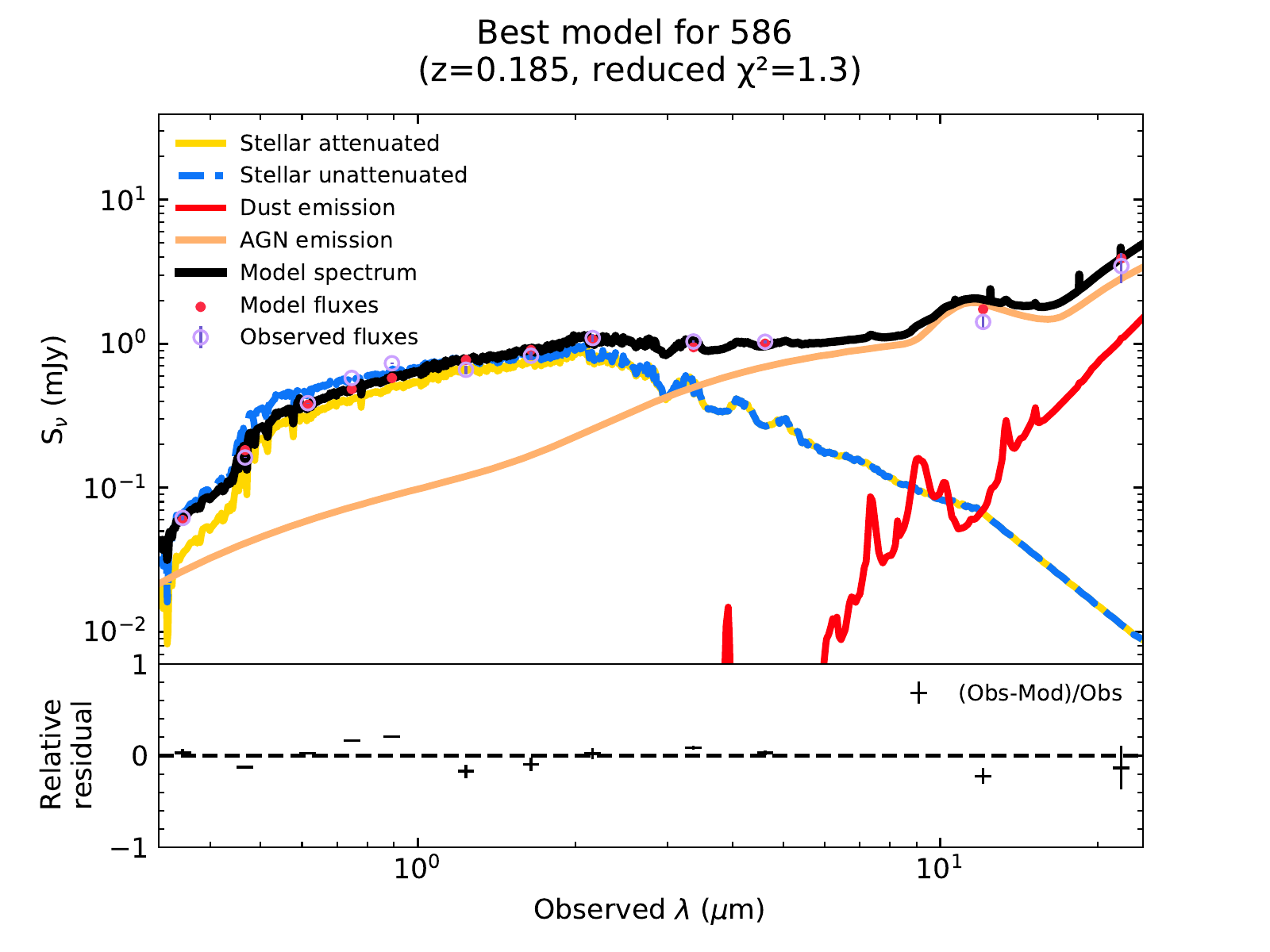}
  \label{}
\end{subfigure}
\begin{subfigure}{.48\textwidth}
  \centering
  \includegraphics[width=1.0\linewidth, height=7.cm]{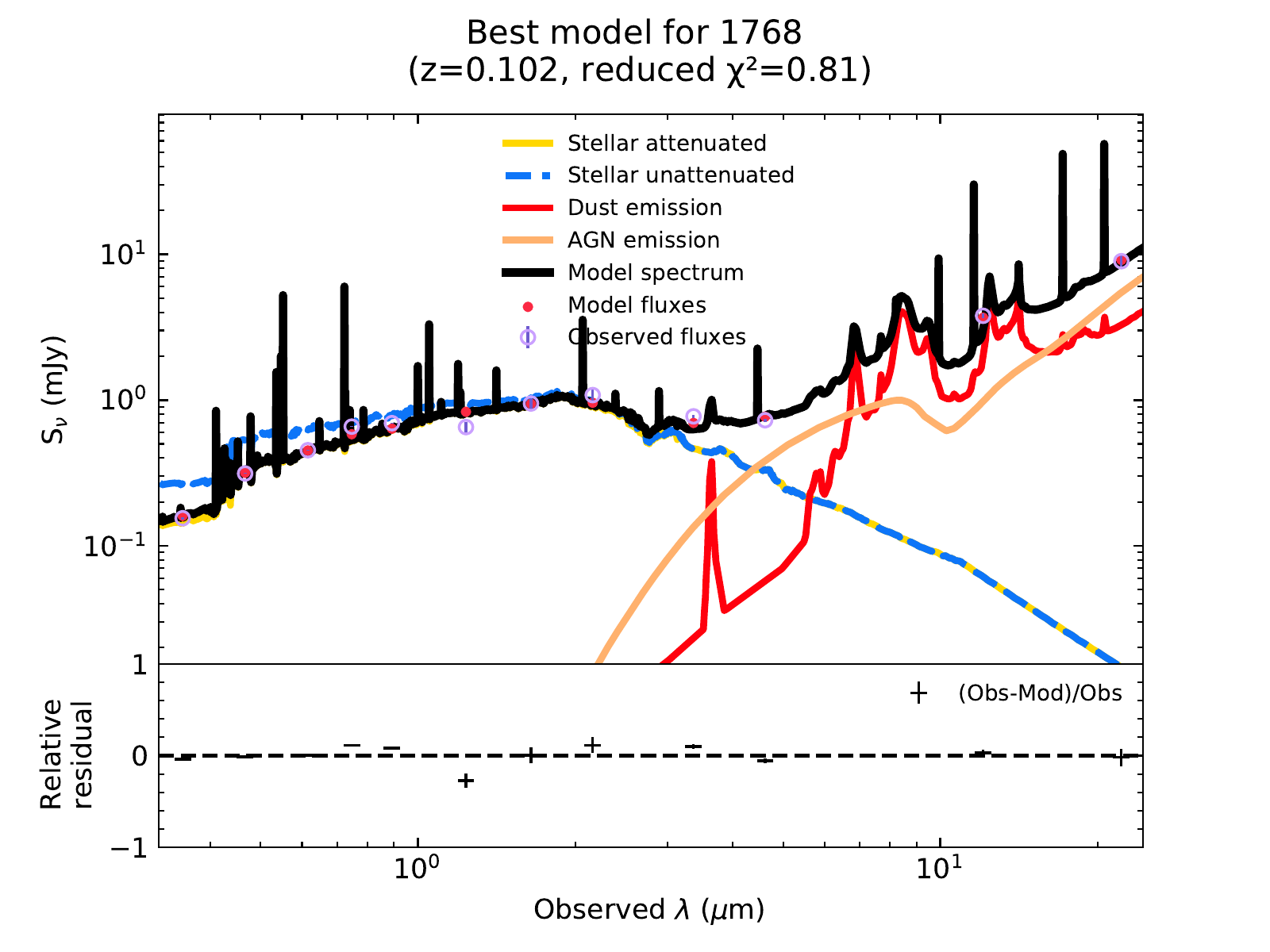}
  \label{}
\end{subfigure}
\begin{subfigure}{.48\textwidth}
  \centering
  \includegraphics[width=1.0\linewidth, height=7.cm]{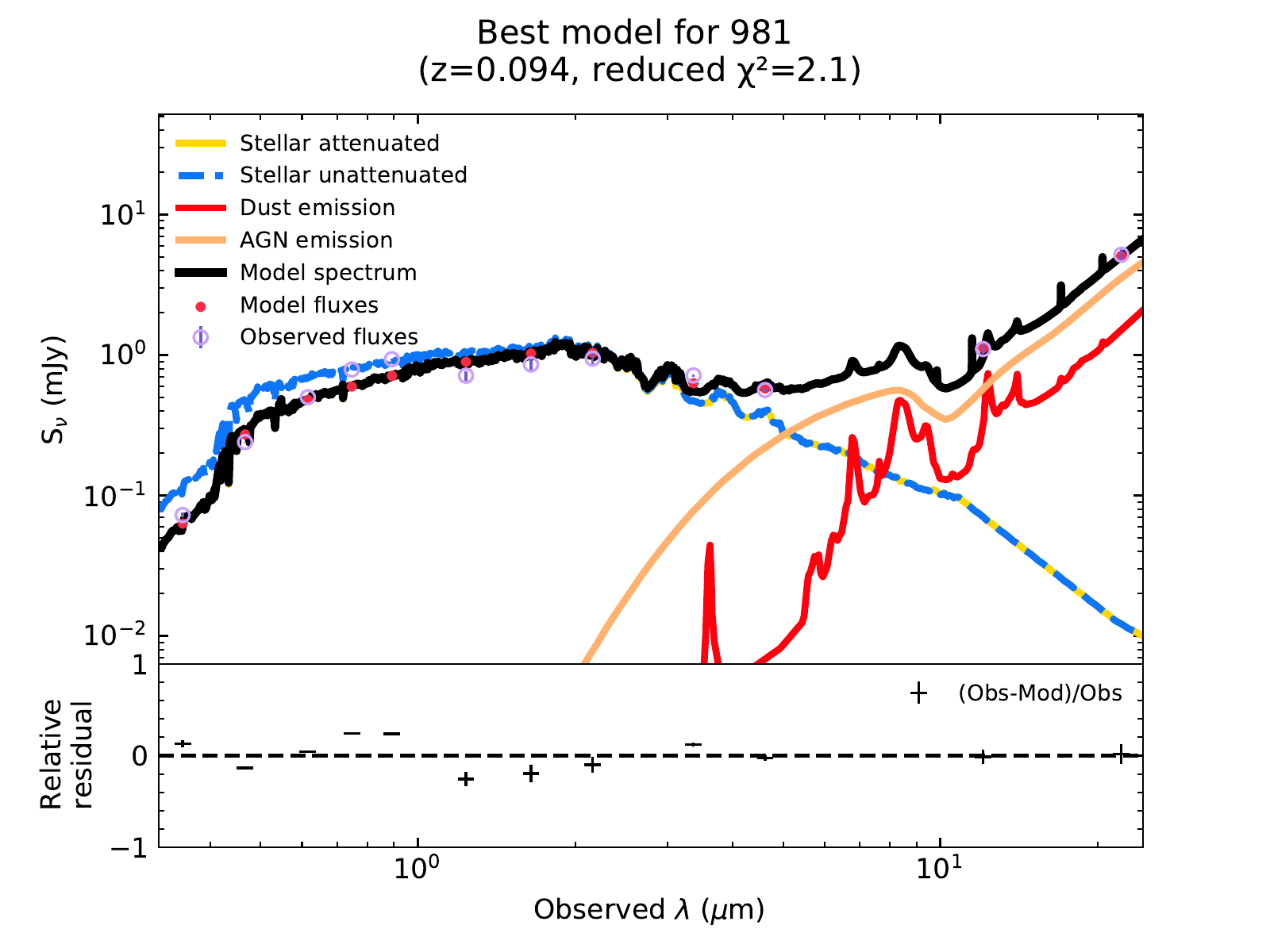}
  \label{}
\end{subfigure}
\caption{Examples of SEDs from sources used in our analysis. Top panels present AGN classified as type 1, based on their optical spectra. Their SFR and stellar mass are $\rm log\,[SFR (M_\odot yr^{-1})]=1.63$,  $\rm log\,[M_*(M_\odot)]=10.06$ (top, left panel) and $\rm log\,[SFR (M_\odot yr^{-1})]=-0.45$ and $\rm log\,[M_*(M_\odot)]=10.75$ (top, right panel), respectively. Bottom panel, presents the SEDs of two AGN classified as type 2, based on their optical spectra. Their SFR and stellar mass are $\rm log\,[SFR (M_\odot yr^{-1})]=1.87$,  $\rm log\,[M_*(M_\odot)]=10.30$ (bottom, left panel) and $\rm log\,[SFR (M_\odot yr^{-1})]=-1.0$ and $\rm log\,[M_*(M_\odot)]=10.24$ (bottom, right panel), respectively.}
\label{fig_SEDs}
\end{figure*}

\begin{figure}
\centering
  \centering
  \includegraphics[width=.92\linewidth, height=10cm]{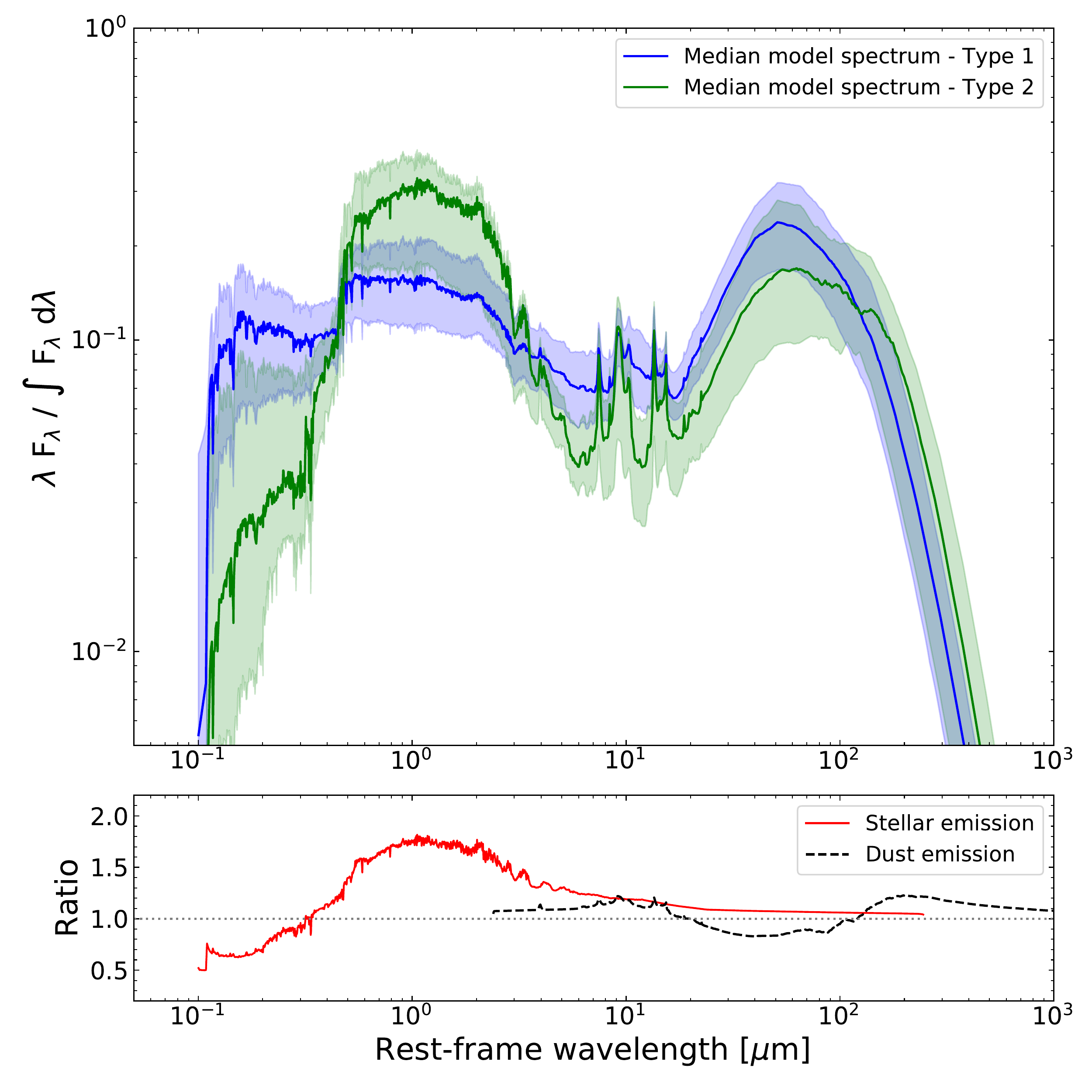}
  \caption{Stacked SEDs at rest-frame for the AGN samples of type 1 (blue) and type 2 (green). The median SEDs of the total emission (host galaxy and AGN components) are plotted with solid lines, while the shaded areas correspond to 15$^{th}$ up to 85$^{th}$ percentiles at each wavelength. The lower panel shows the ratio of the host dust and stellar attenuated emission of the two populations.} 
  \label{fig_stacked}
\end{figure}

\begin{figure}
\centering
  \centering
  \includegraphics[width=.92\linewidth, height=6.2cm]{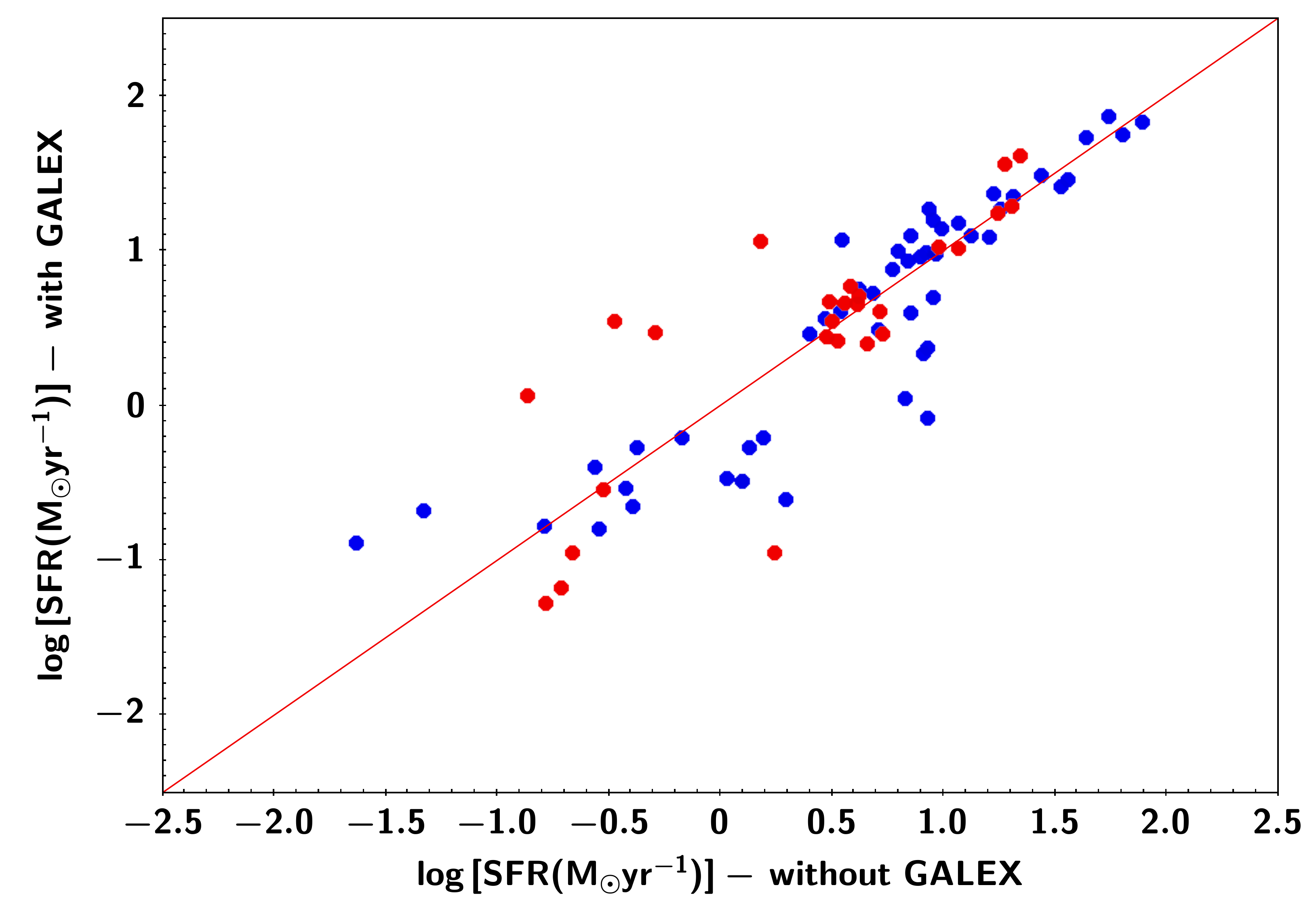}
  \caption{Comparison of SFR measurements with and without GALEX photometry, for the 89 X-ray AGN that have available FUV and NUV and $\chi ^2_{red}<5$ in both runs. Solid line presents the 1:1 relation. Results show that there is no systematic offset between the two measurements. The mean absolute difference of the SFR estimates is 0.25\,dex for type 1 (blue circles) and 0.31\,dex for type 2 AGN (red circles).} 
  \label{fig_galex}
\end{figure}

\subsection{Reliability of SFR measurements}
\label{sec_galex}

SFR is calculated using the emission of young stars that emit most of their light in the UV \citep[e.g.][]{Schreiber2015}. Most of this light is absorbed by interstellar dust and is then re-emitted in the IR part of the spectrum. \cite{Mountrichas2021a, Mountrichas2021c} showed, using X-ray samples from the Bo$\rm \ddot{o}$tes and XMM-{\it{XXL}} fields, that absense of far-IR photometry ({\it{Herschel}}) does not affect the SFR measurements. At high redshifts ($\rm z>0.5$), the $u$ optical band is redshifted to rest-frame wavelengths $<2000\,\mathring{A}$, where radiation from young stars is emitted. However, at lower redshifts shorter wavelengths are required.

\begin{figure}
\centering
  \centering
  \includegraphics[width=.92\linewidth, height=6.2cm]{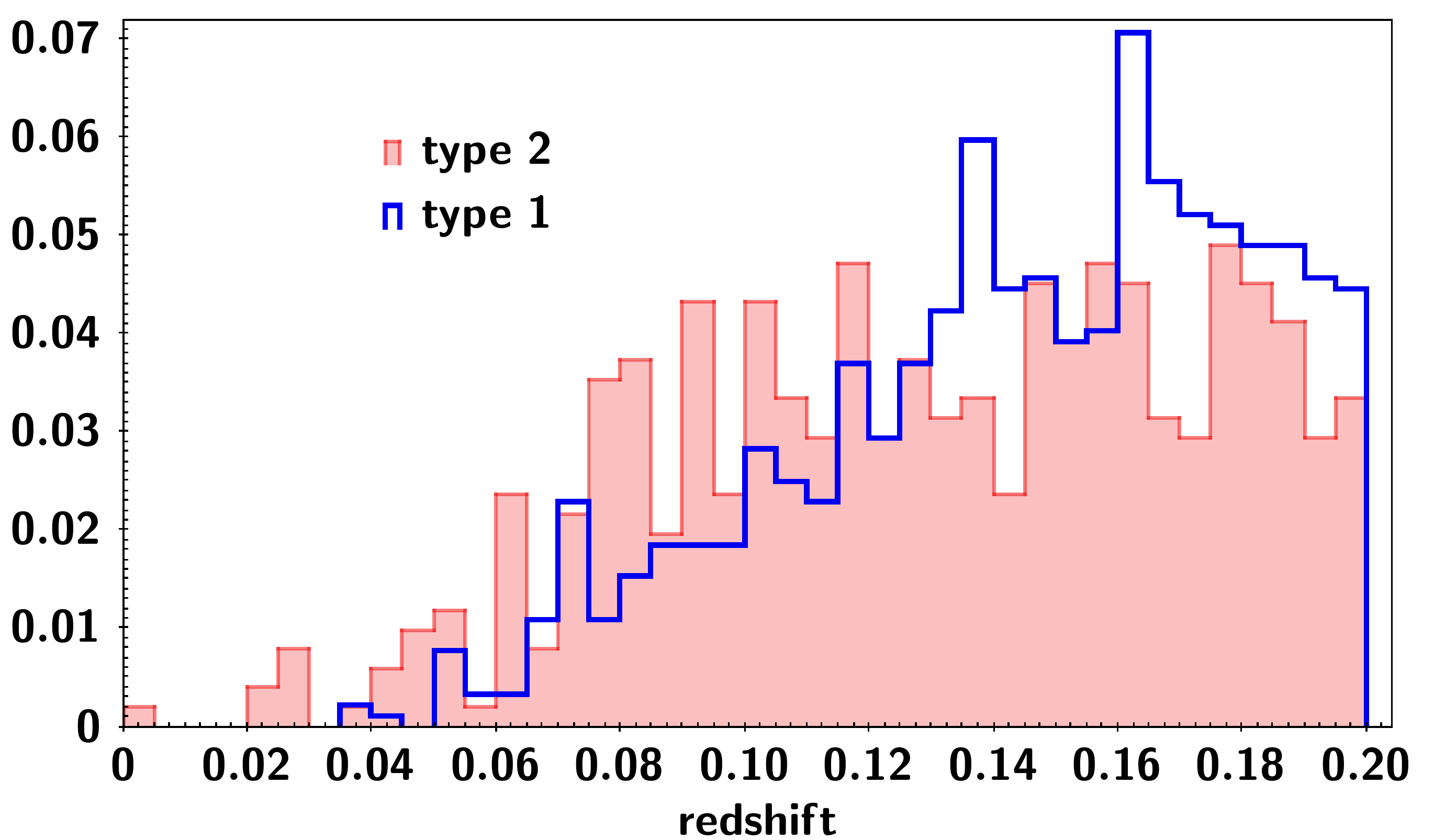}
  \caption{Redshift distributions of type 1 (blue histogram) and type 2 (red histogram) AGN.}
  \label{fig_redz}
\end{figure}

\begin{figure}
\centering
  \centering
  \includegraphics[width=.92\linewidth, height=6.2cm]{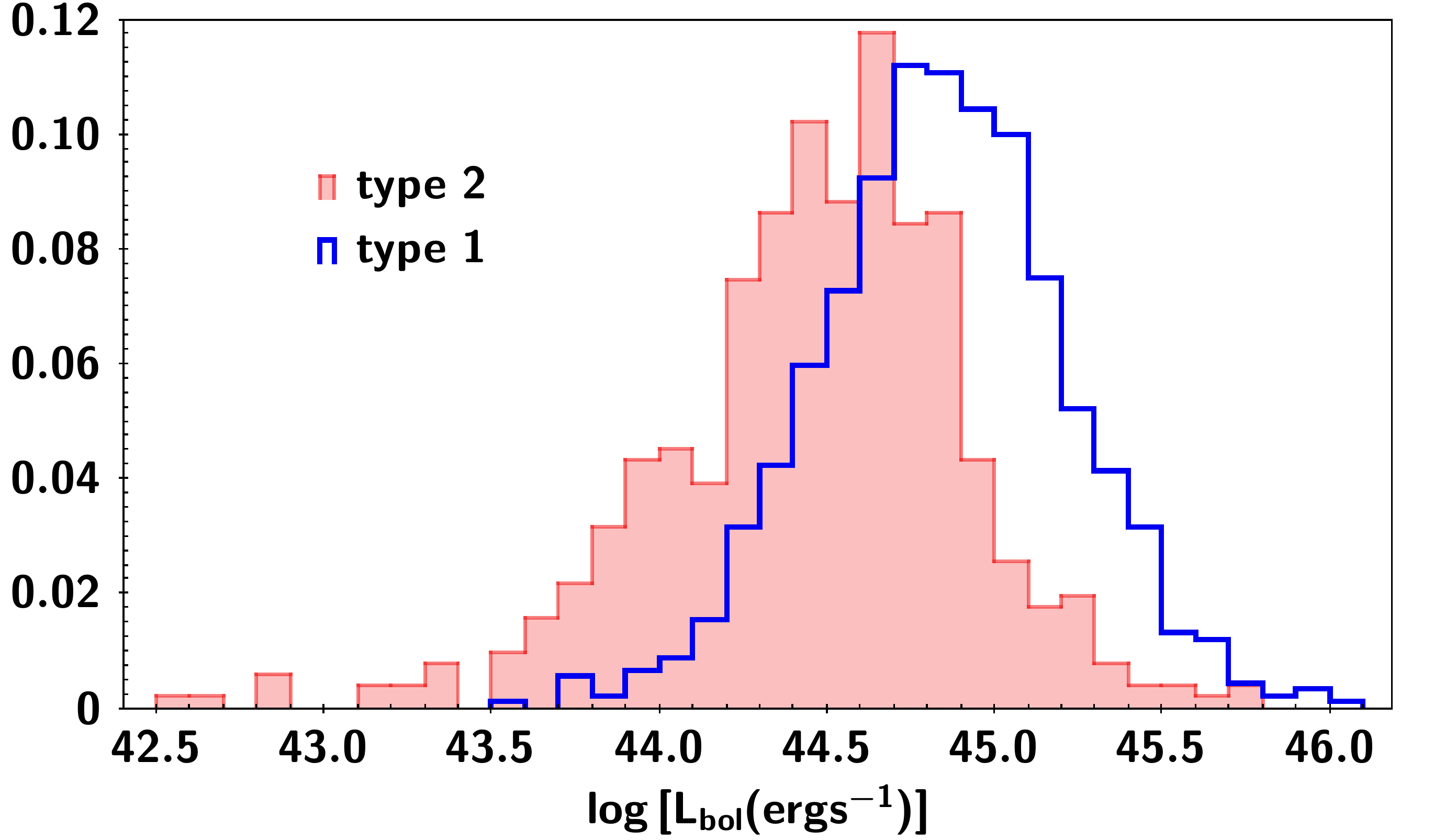}
  \caption{Distributions of bolometric luminosity, $\rm L_{bol}$, of type 1 (blue histogram) and type 2 (red histogram) AGN. $\rm L_{bol}$ has been calculated using the AGN, stellar and dust luminosities, measured by the SED fitting process.}
  \label{fig_lbol}
\end{figure}

To examine whether the absence of UV photometry affects our SFR measurements, we cross match our X-ray sample with the GALEX dataset. There are 114 X-ray AGN that have UV counterparts within a radius of 1$''$. For these sources, we perform SED analysis with and without the far-UV (FUV) and near-UV (NUV) bands of GALEX, using the same parameter space in the fitting process. Fig. \ref{fig_galex} presents the comparison of the SFR measurements, for the 89 sources that have $\chi ^2_{red}<5$ in both runs. Although the sample is small, our results indicate that the two measurements are consistent. The scatter appears larger at SFR<0, but there is no systematic offset between the two calculations. The mean absolute difference of the SFR estimates is 0.25\,dex for type 1 (blue circles) and 0.31\,dex for type 2 AGN (red circles).

\section{Results}
\label{sec_results}

In this section, we compare the SFR and stellar mass of the two AGN types. For the calculation of these two host galaxy properties, we apply SED fitting using the grid described in section \ref{sec_cigale}. Figures \ref{fig_redz} and \ref{fig_lbol}, present the redshift and bolometric luminosity, $\rm L_{bol}$,  distributions of the two AGN populations. Bolometric luminosities have been calculated using the AGN, stellar and dust luminosities, measured by the SED fitting process. To compare the host galaxy properties of the two subsamples, we first account for their different redshift and luminosity distributions. To this end, we assign a weight to each source following the procedure described below. The redshift distributions of the two AGN types, are joined and normalised to the total number of sources in each redshift bin (in bins of 0.1). We repeat the same process for the luminosity distributions (in bins of 0.1\,dex). This effectively gives us the probability density function (PDF) of each source in this 2$-$D (L, z) parameter space. Then, each source is weighted, based on its luminosity and redshift,  $w_{L,\rm z}$, according to the estimated PDF \citep[e.g.,][]{Mountrichas2019, Mountrichas2021a, Mountrichas2021c}. We also use an additional weight that accounts for the uncertainties of the stellar mass and SFR calculations. Specifically, we calculate the significance ($sigma=\rm value/uncertainty$) of each stellar mass, $sigma _{\rm M_*}$, and SFR, $sigma _{\rm SFR}$, measurement and weigh each source based on these values, in addition to the weight that accounts for the redshift and luminosity of each source \citep{Masoura2021}. Thus, the total weight, $w_{\rm t}$, assigned on each source is given by the equation:

\begin{equation}
w_{\rm t}= w_{L,\rm z}\times sigma _{\rm M_*} \times sigma _{\rm SFR}.
\end{equation}

Fig. \ref{fig_mstar_comp} presents the stellar mass distributions of type 1 and 2 X-ray AGN, for the 944 sources that meet our selection criteria (Section \ref{sec_bad_fits}). We notice that the M$_*$ of type 2 sources (red histogram) is shifted to slightly higher values compared to that of type 1 AGN (blue line histogram). The median stellar mass of type 1 AGN is $\rm log\,[M_*(M_\odot)]=10.23^{+0.05}_{-0.08}$, while that of type 2 is $10.49^{+0.16}_{-0.10}\,\rm M_\odot$. Errors have been calculated using bootstrap resampling \citep[e.g.][]{Loh2008}. The stacked SEDs presented in Fig. \ref{fig_stacked}, also reveal that type 2 AGN (green line) have higher stellar mass than their type 1 (blue line) counterparts. Although the result is marginally significant ($\sim 2\,\sigma$), it is in agreement with similar studies at higher redshifts. \cite{Mountrichas2021c} used X-ray AGN in the XMM-{\it{XXL}} field at $\rm 0.1<z<0.9$ and found that type 2 AGN tend to live in host galaxies with stellar mass higher by $\sim 0.3$\,dex compared to their type 1 counterparts. Their classification is based on optical spectra \citep{Menzel2016}. \cite{Zou2019} classified AGN in the COSMOS field, using optical spectra, morphology and optical variability. They found that type 2 AGN prefer to reside in more massive galaxies than type 1, at all redshift spanned by their sample (see their Fig. 6).

In Fig. \ref{fig_sfr_comp}, we present the SFR distributions of type 1 and 2 X-ray AGN, for the 860 sources that meet our selection criteria (Section \ref{sec_bad_fits}). We notice that the distributions of both AGN populations peak at similar SFR values, $\rm log\,[SFR (M_\odot yr^{-1})] \sim 0.7-0.8$. However, type 2 AGN present a second, lower peak at $\rm log\,[SFR (M_\odot yr^{-1})] \sim -0.5$. The SFR distribution of type 1 AGN also has a tail that extends to low SFR values, but this tail is not as prominent as in the case of type 2 AGN. Fig. \ref{fig_stacked} shows that, overall, type 1 AGN have higher star formation than the type 2 population, which can be attributed to the large tail that the SFR distribution of type 2 AGN presents at low values. Similar studies at higher redshifts \citep{Zou2019, Mountrichas2021c}, find that both AGN types have similar SFR distributions and these distributions do not extend to low SFR values. However, \cite{Shimizu2015} used ultra-hard X-ray selected AGN from the $\it{Swift}$ Burst Alert Telescope (BAT), at $\rm z<0.05$. They found that a large fraction of their type 1 and 2 AGN lies below the MS. 

\begin{figure}
\centering
  \centering
  \includegraphics[width=.92\linewidth, height=6.2cm]{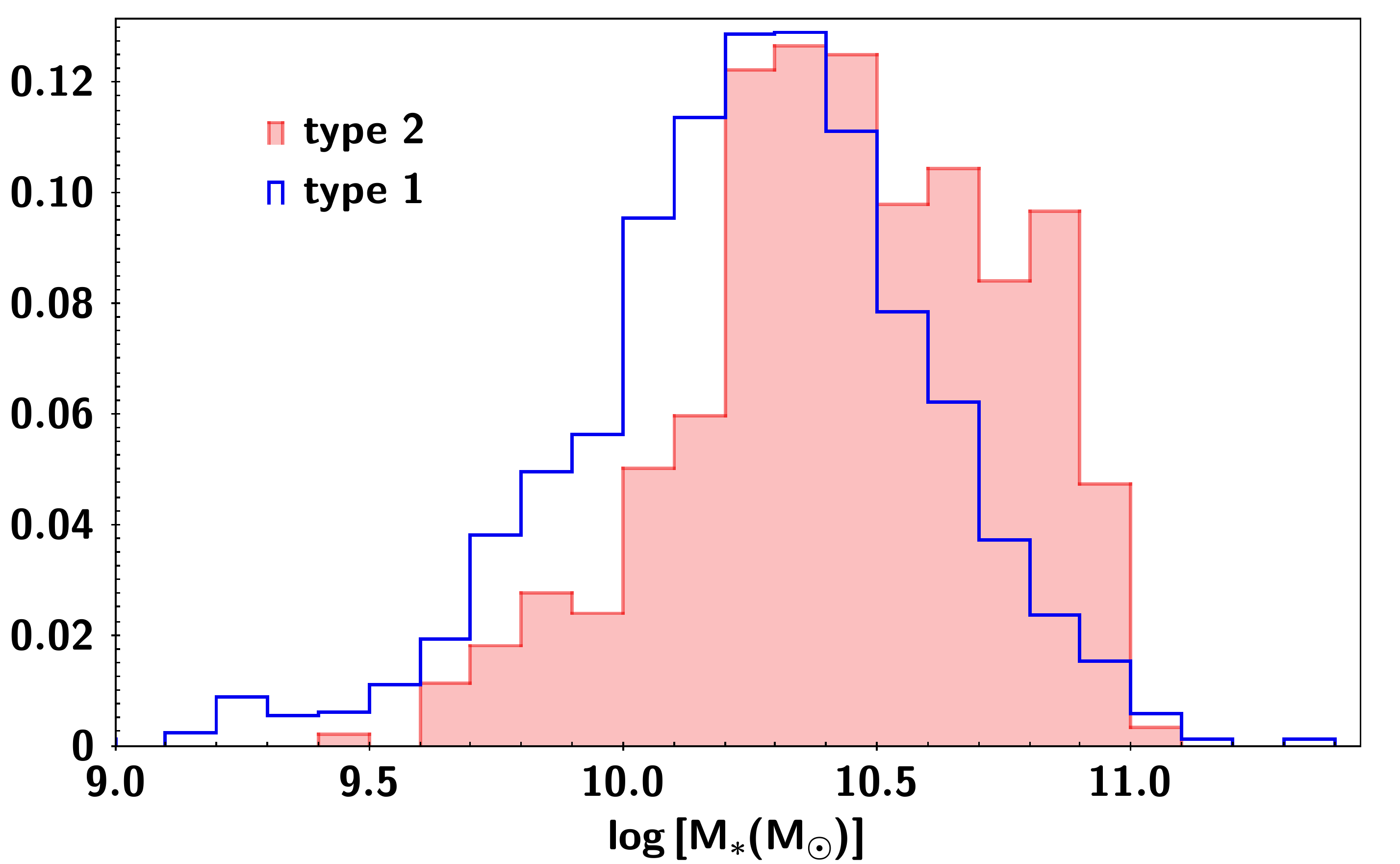}
  \caption{Stellar mass distributions of type 1 (blue line histogram) and type 2 (red shaded histogram) X-ray AGN. Type 2 sources tend to reside in more massive systems compared to their type 1 counterparts. The median stellar mass of type 1 AGN is $\rm log\,[M_*(M_\odot)]=10.23^{+0.05}_{-0.08}$, while that of type 2 is $10.49^{+0.16}_{-0.10}\,\rm M_\odot$. Errors have been calculated using bootstrap resampling.}
  \label{fig_mstar_comp}
\end{figure}%

\begin{figure}
  \centering
  \includegraphics[width=.92\linewidth, height=6.2cm]{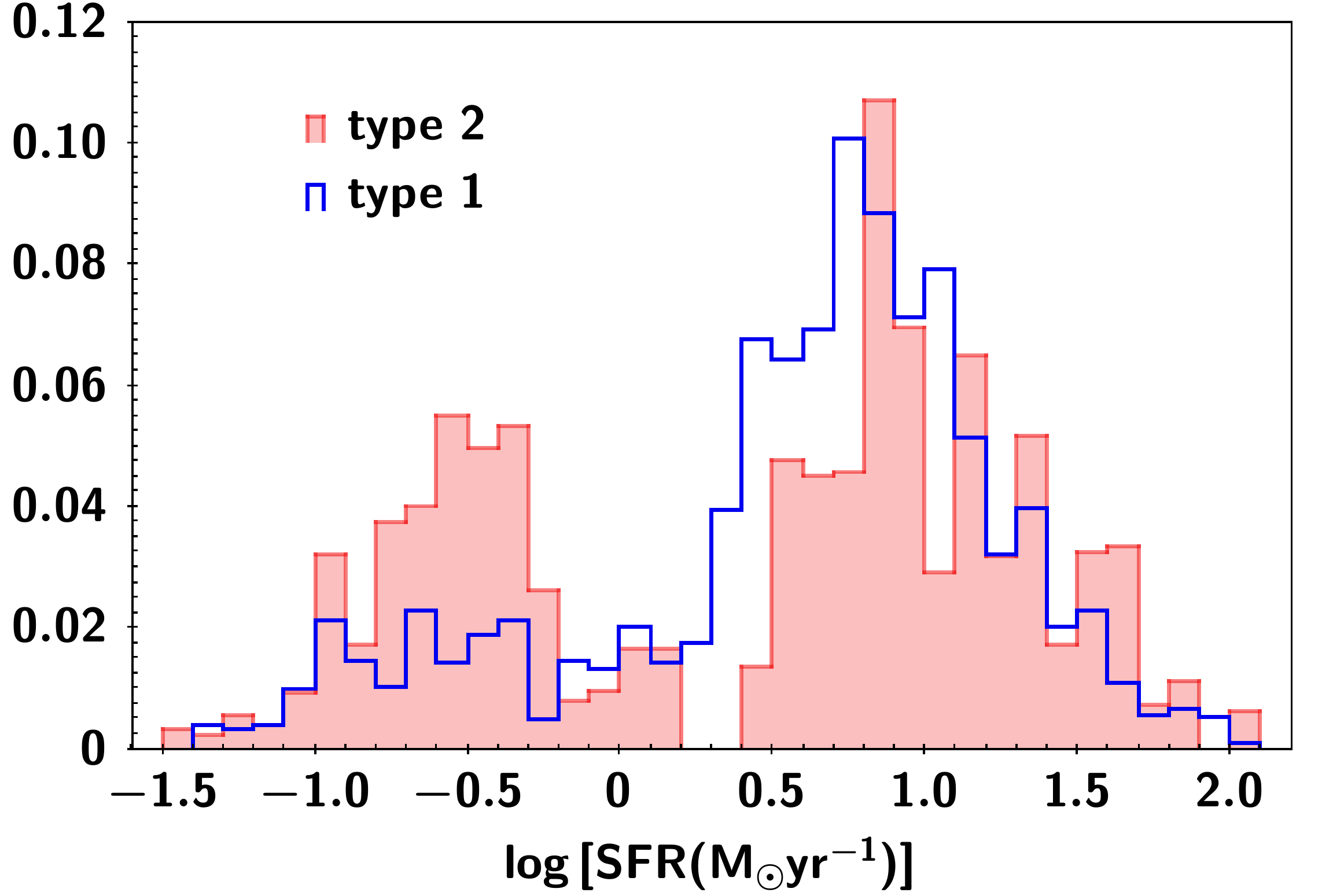}
  \caption{SFR distributions of type 1 (blue line histogram) and type 2 (red histogram) AGN.  Distributions of both AGN populations peak at similar SFR values, $\rm log\,[SFR (M_\odot yr^{-1})] \sim 0.7-0.8$. However, type 2 AGN present a second, lower peak at $\rm log\,[SFR (M_\odot yr^{-1})] \sim -0.5$. The SFR distribution of type 1 AGN also has a tail that extends to low SFR values, but this tail is not as prominent as in the case of type 2 AGN.}
\label{fig_sfr_comp}
\end{figure}

\begin{figure}
  \centering
  \includegraphics[width=.92\linewidth, height=6.2cm]{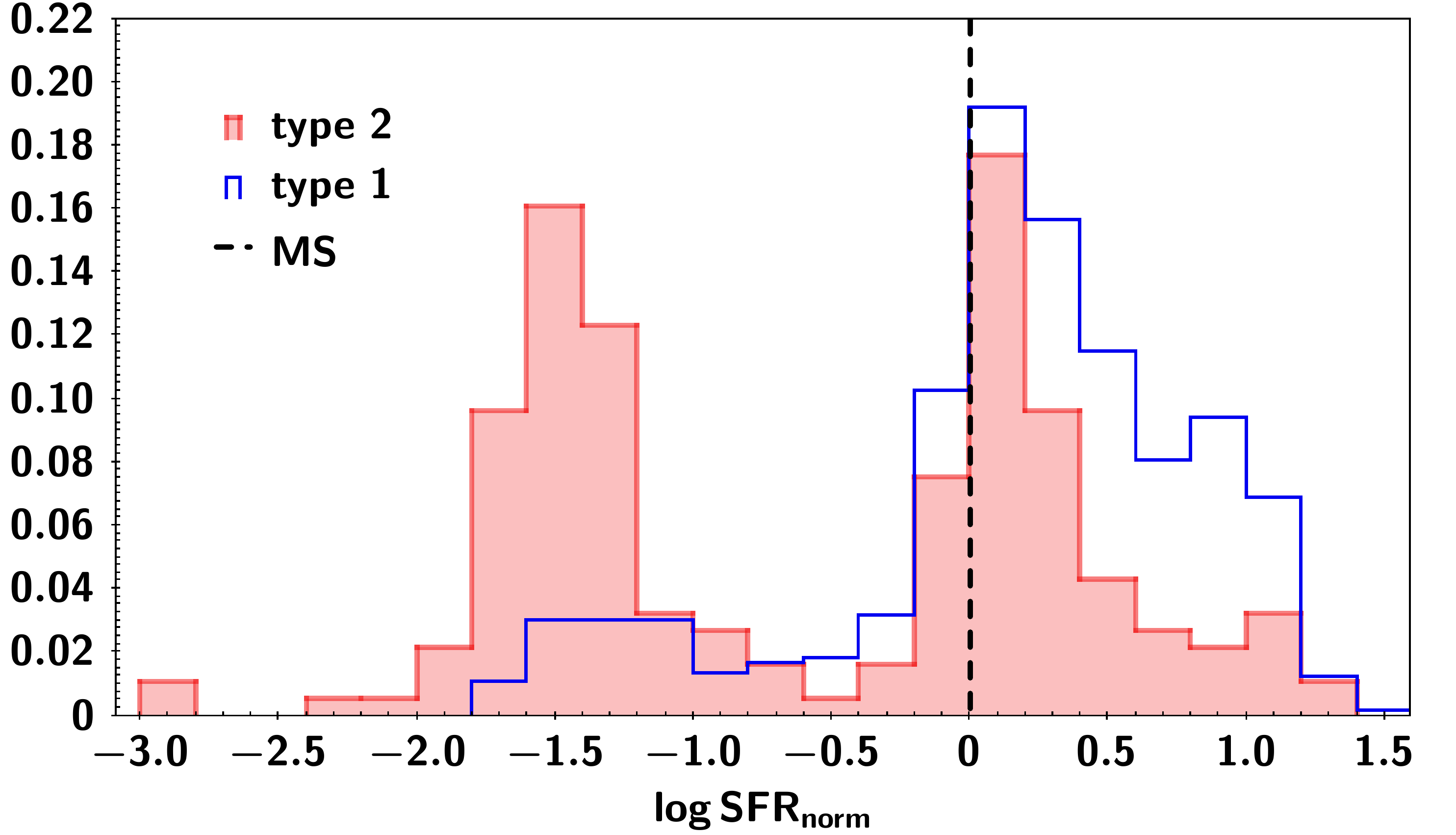}
  \label{}
\caption{Distributions of the SFR$_{norm}$ of type 1 (blue line) and type 2 (red shaded histogram) AGN. Our results show that, in the local universe ($\rm z<0.2$), there is a significant fraction of X-ray AGN that live in quiescent systems (SFR$_{norm}<0$). This appears to be more prominent in the case of type 2 AGN.}
\label{fig_sfrnorm_comp}
\end{figure}

\begin{table*}
\caption{Numbers and fractions of type 1 and 2 AGN that live inside, below and above the MS.}
\centering
\setlength{\tabcolsep}{3.mm}
\begin{tabular}{cccc}
& Above MS & Inside MS & Below MS\\
\hline
 672 type 1 & 305 (45\%) & 260 (39\%) & 107 (16\%)\\
 187 type 2 & 30 (16\%) & 62 (33\%) & 95 (51\%)\\
 \hline
\label{table_fractions}
\end{tabular}
\tablefoot{We consider that an AGN is inside the MS when its SFR$_{norm}$ value is within $0 \pm 1\,\sigma$, where $1\sigma =0.3$\,dex. AGN with SFR$_{norm}>0.3$ live in star forming galaxies, while AGN with AGN with SFR$_{norm}<-0.3$ reside in quiescent systems}
\end{table*}

To examine the position of the AGN relative to the MS in our sample, we estimate the SFR$_{norm}$ parameter. SFR$_{norm}$ is defined as the ratio of the SFR of AGN to the SFR of star-forming MS galaxies with the same stellar mass and redshift \citep[e.g.][]{Masoura2018, Bernhard2019, Masoura2021}. For the calculation of the latter, we use expression 9 from \cite{Schreiber2015}. In this part of the analysis, we require both the SFR and stellar mass criteria, mentioned in Section \ref{sec_bad_fits}, to be fulfilled. 859 AGN meet these requirements. The SFR$_{norm}$ distributions of type 1 and 2  AGN are shown in Fig. \ref{fig_sfrnorm_comp}. The vertical, dashed line indicates the position of the MS. Based on our results, there is a large fraction of AGN, in particular type 2 AGN, that live in quiescent systems. In Table \ref{table_fractions}, we present the fraction of type 1 and 2 AGN that live inside, below and above the MS. Following a similar approach to that of \cite{Shimizu2015}, we consider that an AGN is inside the MS when its SFR$_{norm}$ value is within $0 \pm 1\,\sigma$, where $1\sigma =0.3$\,dex. AGN with SFR$_{norm}>0.3$ live in star forming galaxies, while AGN with AGN with SFR$_{norm}<-0.3$ reside in quiescent systems. Our results show that about half of type 2 AGN live in quiescent systems. This is in agreement with the results of \cite{Shimizu2015} (see their Table 1). Albeit, Shimizu et al. find a similar fraction of quiescent systems for type 1 AGN. In our case, the fraction of type 1 AGN that are below the MS, is significantly lower compared to type 2. We note, however, that the host galaxy properties have been estimated following different approaches in the two studies. In Shimizu et al., SFR has been estimated using IR luminosities and stellar mass measurements are based on the $g-i$ colour (see their Sections 3 and 4, for more details). Furthermore,  \cite{Shimizu2015} define their own MS, using galaxy samples for which they estimate the SFR and M$_*$, by applying similar methods as for their AGN sample. In this work, we have used a relation from the literature to estimate the SFR of MS galaxies \citep{Schreiber2015}. This approach hints a number of systematics \citep{Mountrichas2021c}. For these reasons, a strict comparison between the two studies cannot be made.

In total, 202 AGN ($24\%$), 107 type 1 and 95 type 2, in our dataset live in quiescent systems. Compared to similar studies at higher redshifts this fraction appears increased. \cite{Mountrichas2021b} found that $\sim 12\%$ of the AGN population has SFR$_{norm}<-0.3$ ($\sim 8\%$ of type 1 and $\sim 14\%$ of type 2, see their Fig. 6). However, the fraction of quiescent host galaxies found in our study is lower than that found in the local universe \citep[z<0.05;][]{Shimizu2015}, It is well known that the fraction of quiescent systems, in the overall (non-AGN) galaxy population, increases with cosmic time, in particular for the more massive sources \citep[e.g.,][]{Fontana2009}. About 10\% of galaxies are quiescent at $\rm z \sim 3-4$. The fraction goes up to $\sim 50\%$ at $\rm z \sim 1.5$ and increases even further in the local universe \citep[e.g.,][]{Sanchez2011, Pandya2017}. Although, the definition of quiescent galaxies is not strict and, as mentioned, our approach to define such systems hints systematics, the fraction of sources that host AGN and are quiescent does not appear to be higher than the fraction of quiescent systems in the overall galaxy population. Thus, our results do not seem to support a picture in which AGN suppress the SFR of their host galaxy. Overall, we conclude that our results corroborate that in the local Universe, the fraction of X-ray AGN that live in quiescent systems is higher than that at higher redshifts. This may be more evident in the case of type 2 AGN. However, the majority of AGN live inside or above the star forming main sequence.

\section{Summary}

We have used data from the DR16-SPIDERS-2XRS catalogue to examine the host galaxy properties of type 1 and 2 X-ray AGN, at $\rm z<0.2$. The classification is based on optical spectra. We use optical, near-IR and mid-IR photometry to construct the SEDs of the sources and we fit these SEDs using the CIGALE code. After applying various selection criteria to exclude sources with non reliable stellar mass and SFR measurements we ended up with $\sim 900$ X-ray sources. 

We take into account the different redshift and luminosity distributions of the two AGN types and compare their stellar mass and SFR. We also account for the uncertainties of the measurements of the two galaxy properties, calculated by CIGALE. 

Our analysis shows that type 2 AGN reside in more massive galaxies compared to their type 1 counterparts. Based on our results, type 1 AGN live in galaxies with stellar mass $\rm log\,[M_*(M_\odot)]=10.23^{+0.05}_{-0.08}$, while type 2 reside in galaxies with $10.49^{+0.16}_{-0.10}\,\rm M_\odot$. This finding is in accordance with similar studies at higher redshifts \citep{Zou2019, Mountrichas2021a}.

To investigate the position of X-ray AGN relative to the main sequence, we calculate the SFR$_{norm}$ parameter. Previous studies \citep{Shimizu2015}, found a large fraction of  AGN living in quiescent galaxies, at $\rm z<0.05$. Our results corroborate that in the local universe there is an increased number of AGN ($24\%$) that live in quiescent systems compared to that at higher redshifts. However, the majority of AGN live either inside or above the star forming main sequence.

\begin{acknowledgements}
Lazaros Koutoulidis (LK) acknowledges support from the State
Scholarship Foundation (IKY).\\
GM acknowledges support by the Agencia Estatal de Investigación, Unidad de Excelencia María de Maeztu, ref. MDM-2017-0765.\\
IG and EP acknowledge financial support by the European Union's Horizon 2020 programme "XMM2ATHENA" under grant agreement No 101004168. The research leading to these results has received funding from the European Union's Horizon 2020 programme under the AHEAD2020 project (grant agreement n. 871158).

\end{acknowledgements}

\bibliography{mybib}{}

\begin{thebibliography}{51}
\expandafter\ifx\csname natexlab\endcsname\relax\def\natexlab#1{#1}\fi

\bibitem[{{Akylas} {et~al.}(2006){Akylas}, {Georgantopoulos}, {Georgakakis},
  {Kitsionas}, \& {Hatziminaoglou}}]{Akylas2006}
{Akylas}, A., {Georgantopoulos}, I., {Georgakakis}, A., {Kitsionas}, S., \&
  {Hatziminaoglou}, E. 2006, A\&A, 459, 693

\bibitem[{Antonucci(1993)}]{Antonucci1993}
Antonucci, R. 1993, Annual Review of Astronomy and Astrophysics, 31, 473

\bibitem[{{Barthelmy} {et~al.}(2005){Barthelmy}, {Barbier}, {Cummings},
  {Fenimore}, {Gehrels}, {Hullinger}, {Krimm}, {Markwardt}, {Palmer},
  {Parsons}, {Sato}, {Suzuki}, {Takahashi}, {Tashiro}, \& {Tueller}}]{Bat2005}
{Barthelmy}, S.~D., {Barbier}, L.~M., {Cummings}, J.~R., {et~al.} 2005, SSR,
  120, 143

\bibitem[{Bernhard {et~al.}(2019)Bernhard, Grimmett, Mullaney, Daddi,
  Tadhunter, \& Jin}]{Bernhard2019}
Bernhard, E., Grimmett, L.~P., Mullaney, J.~R., {et~al.} 2019, Monthly Notices
  of the Royal Astronomical Society: Letters, 483, L52

\bibitem[{{Boller} {et~al.}(2016){Boller}, {Freyberg}, {Tr{\"u}mper}, {Haberl},
  {Voges}, \& {Nandra}}]{Boller2016}
{Boller}, T., {Freyberg}, M.~J., {Tr{\"u}mper}, J., {et~al.} 2016, Astronomy
  {\&} Astrophysics, 588, A103

\bibitem[{Boquien {et~al.}(2019)Boquien, Burgarella, Roehlly, Buat, Ciesla,
  Corre, Inoue, \& Salas}]{Boquien2019}
Boquien, M., Burgarella, D., Roehlly, Y., {et~al.} 2019, Astronomy {\&}
  Astrophysics, 622, A103

\bibitem[{Bruzual \& Charlot(2003)}]{Bruzual_Charlot2003}
Bruzual, G. \& Charlot, S. 2003, MNRAS, 344, 1000

\bibitem[{Buat {et~al.}(2021)Buat, Mountrichas, Yang, Boquien, Roehlly,
  Burgarella, Stalevski, Ciesla, \& Theul{\'{e}}}]{Buat2021}
Buat, V., Mountrichas, G., Yang, G., {et~al.} 2021, A\&A, 654, A93

\bibitem[{Charlot \& Fall(2000)}]{Charlot_Fall_2000}
Charlot, S. \& Fall, S.~M. 2000, ApJ, 539, 718

\bibitem[{Ciotti \& Ostriker(1997)}]{Ciotti1997}
Ciotti, L. \& Ostriker, J.~P. 1997, The Astrophysical Journal, 487, L105

\bibitem[{{Comparat} {et~al.}(2020){Comparat}, {Merloni}, {Dwelly}, {Salvato},
  {Schwope}, {Coffey}, \& {Wolf}}]{Comparat2020}
{Comparat}, J., {Merloni}, A., {Dwelly}, T., {et~al.} 2020, Astronomy {\&}
  Astrophysics, 636, A97

\bibitem[{{Dale} {et~al.}(2014){Dale}, {Helou}, {Magdis}, {Armus},
  {D{\'{\i}}az-Santos}, \& {Shi}}]{Dale2014}
{Dale}, D.~A., {Helou}, G., {Magdis}, G.~E., {et~al.} 2014, ApJ, 784, 83

\bibitem[{{Dwelly} {et~al.}(2017){Dwelly}, {Salvato}, {Merloni}, {Brusa},
  {Buchner}, {Anderson}, {Boller}, {Brandt}, {Budav{\'a}ri}, {Clerc}, {Coffey},
  {Del Moro}, {Georgakakis}, {Green}, {Jin}, {Menzel}, {Myers}, {Nandra},
  {Nichol}, {Ridl}, {Schwope}, \& {Simm}}]{Dwelly2017}
{Dwelly}, T., {Salvato}, M., {Merloni}, A., {et~al.} 2017, MNRAS, 469, 1065

\bibitem[{{Ferrarese} \& {Merritt}(2000)}]{Ferrarese2000}
{Ferrarese}, L. \& {Merritt}, D. 2000, ApJ, 539, 9

\bibitem[{Fontana {et~al.}(2009)Fontana, Santini, Grazian, Pentericci, Fiore,
  Castellano, Giallongo, Menci, Salimbeni, Cristiani, Nonino, \&
  Vanzella}]{Fontana2009}
Fontana, A., Santini, P., Grazian, A., {et~al.} 2009, A\&A, 501, 15

\bibitem[{{Gebhardt} {et~al.}(2000)}]{Gebhardt2000}
{Gebhardt}, K. {et~al.} 2000, ApJ, 543, 5

\bibitem[{{H{\"a}ring} \& {Rix}(2004)}]{Haring2004}
{H{\"a}ring}, N. \& {Rix}, H.-W. 2004, ApJl, 604, L89

\bibitem[{Hasinger(2008)}]{Hasinger2008}
Hasinger, G. 2008, Astronomy {\&} Astrophysics, 490, 905

\bibitem[{{Hickox} \& {Alexander}(2018)}]{Hick2018}
{Hickox}, R.~C. \& {Alexander}, D.~M. 2018, ARAA, 56, 625

\bibitem[{{Hopkins} {et~al.}(2008){Hopkins}, {Hernquist}, {Cox}, \&
  {Keres}}]{Hopkins2008a}
{Hopkins}, P.~F., {Hernquist}, L., {Cox}, T.~J., \& {Keres}, D. 2008, ApJS,
  175, 356

\bibitem[{Hopkins {et~al.}(2006)Hopkins, Hernquist, Cox, Matteo, Robertson, \&
  Springel}]{Hopkins2006}
Hopkins, P.~F., Hernquist, L., Cox, T.~J., {et~al.} 2006, The Astrophysical
  Journal Supplement Series, 163, 1

\bibitem[{{Kauffmann} {et~al.}(2003){Kauffmann}, {Heckman}, {White}, {Charlot},
  {Tremonti}, {Peng}, {Seibert}, {Brinkmann}, {Nichol}, {SubbaRao}, \&
  {York}}]{Kauffmann2003}
{Kauffmann}, G., {Heckman}, T.~M., {White}, S. D.~M., {et~al.} 2003, MNRAS,
  341, 54

\bibitem[{Kormendy \& Ho(2013)}]{Kormendy2013}
Kormendy, J. \& Ho, L.~C. 2013, ARAA, 51, 511

\bibitem[{{Lanzuisi} {et~al.}(2017)}]{Lanzuisi2017}
{Lanzuisi}, G. {et~al.} 2017, A\&A, 602, 13

\bibitem[{Loh(2008)}]{Loh2008}
Loh, J.~M. 2008, ApJ, 681, 726

\bibitem[{Magorrian {et~al.}(1998)}]{Magorrian1998}
Magorrian, J. {et~al.} 1998, AJ, 115, 2285

\bibitem[{Masoura {et~al.}(2021)Masoura, Mountrichas, Georgantopoulos, \&
  Plionis}]{Masoura2021}
Masoura, V.~A., Mountrichas, G., Georgantopoulos, I., \& Plionis, M. 2021,
  Astronomy {\&} Astrophysics, 646, A167

\bibitem[{Masoura {et~al.}(2018)Masoura, Mountrichas, Georgantopoulos, Ruiz,
  Magdis, \& Plionis}]{Masoura2018}
Masoura, V.~A., Mountrichas, G., Georgantopoulos, I., {et~al.} 2018, A\&A, 618,
  31

\bibitem[{Menzel {et~al.}(2016)}]{Menzel2016}
Menzel, M.-L. {et~al.} 2016, MNRAS, 457, 110

\bibitem[{Mountrichas {et~al.}(2021{\natexlab{a}})Mountrichas, Buat,
  Georgantopoulos, Yang, Masoura, Boquien, \& Burgarella}]{Mountrichas2021b}
Mountrichas, G., Buat, V., Georgantopoulos, I., {et~al.} 2021{\natexlab{a}},
  Astronomy {\&} Astrophysics, 653, A70

\bibitem[{Mountrichas {et~al.}(2021{\natexlab{b}})Mountrichas, Buat, Yang,
  Boquien, Burgarella, \& Ciesla}]{Mountrichas2021a}
Mountrichas, G., Buat, V., Yang, G., {et~al.} 2021{\natexlab{b}}, Astronomy
  {\&} Astrophysics, 646, A29

\bibitem[{Mountrichas {et~al.}(2021{\natexlab{c}})Mountrichas, Buat, Yang,
  Boquien, Burgarella, Ciesla, Malek, \& Shirley}]{Mountrichas2021c}
Mountrichas, G., Buat, V., Yang, G., {et~al.} 2021{\natexlab{c}}, Astronomy
  {\&} Astrophysics, 653, A74

\bibitem[{Mountrichas {et~al.}(2019)Mountrichas, Georgakakis, \&
  Georgantopoulos}]{Mountrichas2019}
Mountrichas, G., Georgakakis, A., \& Georgantopoulos, I. 2019, Monthly Notices
  of the Royal Astronomical Society, 483, 1374

\bibitem[{Netzer(2015)}]{Netzer2015}
Netzer, H. 2015, Annual Review of Astronomy and Astrophysics, 53, 365

\bibitem[{Pandya {et~al.}(2017)Pandya, Brennan, Somerville, Choi, Barro, Wuyts,
  Taylor, Behroozi, Kirkpatrick, Faber, Primack, Koo, McIntosh, Kocevski, Bell,
  Dekel, Fang, Ferguson, Grogin, Koekemoer, Lu, Mantha, Mobasher, Newman,
  Pacifici, Papovich, van~der Wel, \& Yesuf}]{Pandya2017}
Pandya, V., Brennan, R., Somerville, R.~S., {et~al.} 2017, A\&A, 472, 2054

\bibitem[{Pouliasis {et~al.}(2020)Pouliasis, Mountrichas, Georgantopoulos,
  Ruiz, Yang, \& Bonanos}]{Pouliasis2020}
Pouliasis, E., Mountrichas, G., Georgantopoulos, I., {et~al.} 2020, Monthly
  Notices of the Royal Astronomical Society, 495, 1853

\bibitem[{Prevot {et~al.}(1984)Prevot, Lequeux, Maurice, Prevot, \&
  Rocca-Volmerange}]{Prevot1984}
Prevot, M., Lequeux, J., Maurice, E., Prevot, L., \& Rocca-Volmerange, B. 1984,
  A\&A, 132, 389

\bibitem[{{Salvato} {et~al.}(2018){Salvato}, {Buchner}, {Budav{\'a}ri},
  {Dwelly}, {Merloni}, {Brusa}, {Rau}, {Fotopoulou}, \& {Nandra}}]{Salvato2018}
{Salvato}, M., {Buchner}, J., {Budav{\'a}ri}, T., {et~al.} 2018, MNRAS, 473,
  4937

\bibitem[{S{\'{a}}nchez {et~al.}(2011)S{\'{a}}nchez, Pozzi, Gruppioni, Cimatti,
  Ilbert, Pozzetti, McCracken, Capak, Floch, Salvato, Zamorani, Carollo,
  Contini, Kneib, F{\`{e}}vre, Lilly, Mainieri, Renzini, Scodeggio, Bardelli,
  Bolzonella, Bongiorno, Caputi, Coppa, Cucciati, de~la Torre, de~Ravel,
  Franzetti, Garilli, Iovino, Kampczyk, Knobel, Kova{\v{c}}, Lamareille,
  Borgne, Brun, Maier, Mignoli, Pell{\'{o}}, Peng, Perez-Montero, Ricciardelli,
  Silverman, Tanaka, Tasca, Tresse, Vergani, \& Zucca}]{Sanchez2011}
S{\'{a}}nchez, H.~D., Pozzi, F., Gruppioni, C., {et~al.} 2011, A\&A, 417, 900

\bibitem[{Schreiber {et~al.}(2015)}]{Schreiber2015}
Schreiber, C. {et~al.} 2015, A\&A, 575, 29

\bibitem[{Shimizu {et~al.}(2015)Shimizu, Mushotzky, Mel{\'{e}}ndez, Koss, \&
  Rosario}]{Shimizu2015}
Shimizu, T.~T., Mushotzky, R.~F., Mel{\'{e}}ndez, M., Koss, M., \& Rosario,
  D.~J. 2015, Monthly Notices of the Royal Astronomical Society, 452, 1841

\bibitem[{{Skrutskie} {et~al.}(2006){Skrutskie}, {Cutri}, {Stiening},
  {Weinberg}, {Schneider}, {Carpenter}, {Beichman}, {Capps}, {Chester},
  {Elias}, {Huchra}, {Liebert}, {Lonsdale}, {Monet}, {Price}, {Seitzer},
  {Jarrett}, {Kirkpatrick}, {Gizis}, {Howard}, {Evans}, {Fowler}, {Fullmer},
  {Hurt}, {Light}, {Kopan}, {Marsh}, {McCallon}, {Tam}, {Van Dyk}, \&
  {Wheelock}}]{Mass2006}
{Skrutskie}, M.~F., {Cutri}, R.~M., {Stiening}, R., {et~al.} 2006, AJ, 131,
  1163

\bibitem[{{Somerville} {et~al.}(2008){Somerville}, {Hopkins}, J., {Robertson},
  \& L.}]{Somerville2008}
{Somerville}, R.~S., {Hopkins}, P.~F., J., C.~T., {Robertson}, B.~E., \& L., H.
  2008, MNRAS, 391, 481

\bibitem[{Stalevski {et~al.}(2012)Stalevski, Fritz, Baes, Nakos, \&
  Popovi{\'{c}}}]{Stalevski2012}
Stalevski, M., Fritz, J., Baes, M., Nakos, T., \& Popovi{\'{c}}, L.~{\v{C}}.
  2012, Monthly Notices of the Royal Astronomical Society, 420, 2756

\bibitem[{Stalevski {et~al.}(2016)Stalevski, Ricci, Ueda, Lira, Fritz, \&
  Baes}]{Stalevski2016}
Stalevski, M., Ricci, C., Ueda, Y., {et~al.} 2016, Monthly Notices of the Royal
  Astronomical Society, 458, 2288

\bibitem[{{Suh} {et~al.}(2019){Suh}, {Civano}, {Hasinger}, {Lusso}, {Marchesi},
  {Schulze}, {Onodera}, {Rosario}, \& {Sanders}}]{Suh2019}
{Suh}, H., {Civano}, F., {Hasinger}, G., {et~al.} 2019, ApJ, 872, 168

\bibitem[{{Ueda} {et~al.}(2014){Ueda}, {Akiyama}, {Hasinger}, {Miyaji}, \&
  {Watson}}]{Ueda2014}
{Ueda}, Y., {Akiyama}, M., {Hasinger}, G., {Miyaji}, T., \& {Watson}, M.~G.
  2014, ApJ, 786, 104

\bibitem[{Urry \& Padovani(1995)}]{Urry1995}
Urry, C.~M. \& Padovani, P. 1995, Publications of the Astronomical Society of
  the Pacific, 107, 803

\bibitem[{{Wright} {et~al.}(2010){Wright}, {Eisenhardt}, {Mainzer}, {Ressler},
  {Cutri}, {Jarrett}, {Kirkpatrick}, {Padgett}, {McMillan}, {Skrutskie},
  {Stanford}, {Cohen}, {Walker}, {Mather}, {Leisawitz}, {Gautier}, {McLean},
  {Benford}, {Lonsdale}, {Blain}, {Mendez}, {Irace}, {Duval}, {Liu}, {Royer},
  {Heinrichsen}, {Howard}, {Shannon}, {Kendall}, {Walsh}, {Larsen}, {Cardon},
  {Schick}, {Schwalm}, {Abid}, {Fabinsky}, {Naes}, \& {Tsai}}]{Wright2010}
{Wright}, E.~L., {Eisenhardt}, P.~R.~M., {Mainzer}, A.~K., {et~al.} 2010, AJ,
  140, 1868

\bibitem[{Yang {et~al.}(2020)Yang, Boquien, Buat, Burgarella, Ciesla, Duras,
  Stalevski, Brandt, \& Papovich}]{Yang2020}
Yang, G., Boquien, M., Buat, V., {et~al.} 2020, Monthly Notices of the Royal
  Astronomical Society, 491, 740

\bibitem[{Zou {et~al.}(2019)Zou, Yang, Brandt, \& Xue}]{Zou2019}
Zou, F., Yang, G., Brandt, W.~N., \& Xue, Y. 2019, The Astrophysical Journal,
  878, 11

\end{thebibliography}
\bibliographystyle{aa}

\end{document}